\def\year{2021}\relax
\documentclass[letterpaper]{article} % DO NOT CHANGE THIS
\usepackage{aaai21}  % DO NOT CHANGE THIS
\usepackage{times}  % DO NOT CHANGE THIS
\usepackage{helvet} % DO NOT CHANGE THIS
\usepackage{courier}  % DO NOT CHANGE THIS
\usepackage[hyphens]{url}  % DO NOT CHANGE THIS
\usepackage{graphicx} % DO NOT CHANGE THIS
\usepackage{natbib}  % DO NOT CHANGE THIS AND DO NOT ADD ANY OPTIONS TO IT
\usepackage{caption} % DO NOT CHANGE THIS AND DO NOT ADD ANY OPTIONS TO IT
\usepackage[switch]{lineno}
\usepackage{subcaption}
\usepackage{multirow}
\usepackage{adjustbox}
\usepackage{amssymb}
\usepackage{amsmath}
\usepackage[noend]{algorithmic}
\usepackage[ruled,vlined]{algorithm2e}
\usepackage{xcolor}

\newcommand{\npos}{{\rho}}
\newcommand{\nneg}{{\eta}}
\newcommand{\recall}[1]{\text{Recall@}#1}
\newcommand{\uv}{\alpha}
\newcommand{\iv}{\beta}

\newcommand{\precision}[1]{\text{Precision@}#1}
\newcommand{\map}[1]{\text{MAP@}#1}
\newcommand{\ndcg}[1]{\text{NDCG@}#1}

\newcommand{\indicator}{\mathbb{I}}
\newcommand{\argmax}{\text{argmax}}
\newcommand{\softmax}{\text{softmax}}
\newcommand{\perm}[1]{P^{(#1)}}
\newcommand{\hperm}[1]{{\tilde{P}^{(#1)}}}
\newcommand{\onecol}{\mathbf{1}}
\newcommand{\ourmodel}[1]{$\text{DRM}_{\text{#1}}$}

\newcommand{\loss}{\mathcal{L}}
\newcommand{\hingeloss}{\loss_{\text{hinge}}}
\newcommand{\neuloss}{\loss_{\text{DRM}}}
\newcommand{\ourloss}{\loss}
\newcommand{\uset}{\mathcal{U}}
\newcommand{\iset}{\mathcal{V}}

\newcommand{\obj}{\mathcal{O}}

\newcommand{\hit}{\text{Hit}}

\title{A Differentiable Ranking Metric Using Relaxed \\ Sorting Operation for Top-K Recommender Systems}
\author {
    Hyunsung Lee\textsuperscript{\rm 1}, Yeongjae Jang\textsuperscript{\rm 2}, Jaekwang Kim\textsuperscript{\rm 3} Honguk Woo\textsuperscript{\rm 4}\\
}
\affiliations {
    \textsuperscript{\rm 1} Department of Electrical and Computer Engineering, Sungkyunkwan University, South Korea \\ 
    \textsuperscript{\rm 2} Department of Mathematics, Sungkyunkwan University, South Korea \\
    \textsuperscript{\rm 3} Institute for Convergence, Sungkyunkwan University, South Korea \\
    \textsuperscript{\rm 4} Department of Computer Science and Engineering, Sungkyunkwan University, South Korea
    
    \{hyunsung.lee, ja3156, linux, hwoo\}@skku.edu 
    
}

\begin{document}
\maketitle

\begin{abstract}
%200 words is a nice upper bound of abstract we should follow;
%80 words is a nice lower bound of abstract we should follow;
A recommender system generates personalized recommendations for a user by computing the preference score of items, sorting the items according to the score, and filtering top-$K$ items with high scores.  
While sorting and ranking items are integral for this recommendation procedure, it is nontrivial to incorporate them in the process of end-to-end model training since sorting is nondifferentiable and hard to optimize with gradient descent. 
This incurs the inconsistency issue between existing learning objectives and ranking metrics of recommenders. 
In this work, we present DRM (differentiable ranking metric) that mitigates the inconsistency and improves recommendation performance by employing the differentiable relaxation of ranking metrics. 
Via experiments with several real-world datasets, we demonstrate that the joint learning of the DRM objective upon existing factor based recommenders significantly improves the quality of recommendations, in comparison with other state-of-the-art recommendation methods.
\end{abstract}

\section{Introduction}
With the massive growth of online content, it has become common for online content platforms to operate recommender systems that provide personalized recommendations, aiming to facilitate better user experiences and alleviate the dilemma of choices \cite{schwartz2018paradox}.  
In general, recommender systems generate the relevance score of items with respect to a user and recommend top-$K$ items of high scores. Thus, sorting (or ranking) items serves an important role in top-$K$ recommendations.

However, model based recommenders usually exploit objectives such as mean squared error or log-likelihood that are limited in accurately reflecting the ranking nature of top-$K$ recommendations. It is because the sorting operation is nondifferentiable, and incorporating it into the end-to-end model training using gradient descent is challenging.
As has been noted in several works~\cite{LambdaRankOuBible,xu2008directly,listwisecollaborativefiltering}, optimizing objectives that do not consider the ranking nature of top-$K$ recommendations does not always guarantee the best performance.
Although there exist ranking-oriented objectives: pairwise objectives such as Bayesian personalized ranking~\cite{BPR} and listwise objectives based on Plackett-Luce distribution~\cite{SQLRANK,listwisecollaborativefiltering}, neither objectives fit well with top-$K$ recommendations. 
Pairwise objectives consider only the  ranking between a pair of items, while listwise objectives consider all the items yet with equal importance regardless of their ranks.

To bridge such inconsistency between the existing learning objectives commonly used for training recommenders and the ranking nature of top-$K$ recommendations, we present DRM (differentiable ranking metric), which is a differentiable relaxation scheme of ranking metrics such as \precision{K} or \recall{K}. By employing the differentiable relaxation scheme for sorting operation~\cite{grover2018stochastic}, DRM expedites direct optimization of the ranking metrics.

% Contribution 1
To do so, we first reformulate the ranking metrics in terms of permutation matrix arithmetic forms, and then relax the nondifferentiable permutation matrix in the arithmetic forms to a differentiable row-stochastic matrix. This reformulation and relaxation allow us to optimize ranking metrics in a differentiable form of DRM. Using DRM as an optimization objective renders end-to-end recommendation model training highly consistent with ranking metrics. Moreover, DRM can be readily incorporated into existing recommenders via joint learning with their own objectives without modifying their structure, preserving the benefits of the models.

To evaluate the effect of DRM upon existing models, we adopt two state-of-the-art factor based recommenders, WARP~\cite{weston2013learning} and CML~\cite{CML}. Our experiments demonstrate that the DRM objective significantly improves
the performance of top-$K$ recommendations on several real-world datasets.

\section{Preliminaries} \label{sec:prelim}
Given a set of $M$ users $\uset = \{1, 2, \dots, M\}$, and a set of $N$ items $\iset = \{1, 2, \dots, N\}$, and a set of interactions $y_{u, i}$ for all users $u$ in $\uset$ and all items $i$ in $\iset$, a recommendation model is learned to predict preference or score $\hat{y}_{u, i} \in \mathbb{R}$ of user $u$ to item $i$. We use predicted preference and predicted score interchangeably to denote $\hat{y}_{u, i}$.
We use binary implicit feedback $y_{u, i}$ such that $y_{u, i} = 1$ if user $u$ has interacted with item $i$, and 0 otherwise.
Note that we only consider this binary feedback format in this work, while our approach can be generalized for other implicit feedback settings. 

For user $u$, we use $i$ to represent positive items that $u$ has interacted with, and $j$ to represent negative items that $u$ has not. 
In addition, we use $\iset_u$ to represent a set of positive items for user $u$, and use ${y}_u$ for its bag-of-words notation, i.e., a column vector $[y_{u, 1}, y_{u, 2}, \dots, y_{u, n}]^T$. 
Similarly, we use $\hat{{y}}_u$ to represent the vector of predicted scores of items, i.e., $[\hat{y}_{u, 1}, \hat{y}_{u, 2}, \dots, \hat{y}_{u, n}]^T$.

\subsection{Objectives for Recommendation Models}\label{ssec:obj}
In general, the objectives of recommenders are categorized into pointwise, pairwise, and listwise.

Pointwise objectives maximize the prediction accuracy independently from the errors of item rankings. 
While several pointwise objectives, such as mean squared error and cross-entropy, are commonly used, those objectives are known to have limitations in that small errors of the objectives do not always lead to high-quality recommendations~\cite{recperformance}.

In the recommender system domain, pairwise objectives have gained popularity because they are more closely related to top-$K$ recommendations than pointwise objectives. 
Model training with pairwise objectives enables a recommender to learn users' preferences by casting the recommendation task as a binary classification, predicting whether user $u$ prefers item $i$ to item $j$. 
For example, Bayesian personalized ranking~\cite{BPR} minimizes negative log-likelihood of the probability that user $u$ prefers item $i$ to item $j$ by 

\begin{equation*}
    \loss_{\text{BPR}} = - \sum_{u \in \uset} \sum_{i \in \iset_u} \sum_{j \in \iset - \iset_u} \log \sigma(\hat{y}_{u, i} - \hat{y}_{u, j})
\end{equation*} where $\sigma(\cdot)$ is a sigmoid function, and $\sigma(\hat{y}_{u, i} - \hat{y}_{u, j})$ is interpreted as the probability of user $u$ prefers item $i$ to item $j$.

Another popular pairwise objective is weighted hinge loss,

\begin{equation} \label{eq:hinge}
    \hingeloss = \sum_{u \in \uset} \sum_{i \in \iset_u} \sum_{j \in \iset - \iset_u} \Phi_{ui} [\mu - \hat{y}_{u, i} + \hat{y}_{u, j}]_{+}
\end{equation} where $[x]_+ = \max(0, x)$, and $\mu$ is a margin value. The weight $\Phi_{ui}$ introduced in~\cite{warp} enables pairwise objectives to emphasize the loss of positive items at lower ranks; the value of $\Phi_{ui}$ is chosen to be larger if the approximated rank  is lower for a positive item $i$. Some choices of $\Phi_{ui}$ are known to make the optimizing hinge loss closely related to maximizing discounted cumulative gain~\cite{lim2015top,liang2018top}.

For top-$K$ recommendations, listwise objectives have been recently explored by a few research works~\cite{cao2007learning,listrank,listwisecollaborativefiltering}. In general, listwise objectives are based on Plackett-Luce probability distribution of list permutations, i.e.,

\begin{equation*}
    \perm{\hat{y}} = \prod_{j=1}^{\min(K, N)} \frac{\phi (\hat{y}_{\pi_j})}{\sum_{l=j}^{N} \phi(\hat{y}_{\pi_l})}
\end{equation*}
where $\phi(\cdot)$ is an arbitrary smoothing function, e.g., $\phi(\cdot) = \exp(\cdot)$. 
These listwise objectives aim to maximize the probability of correctly ordered permutations by minimizing log-likelihood or cross-entropy. However, they have a limitation of high computational complexity in that the complexity of calculating the permutation probability grows exponentially as the number of items in the dataset increases.

%In this work, we consider the ranking nature and weights on top-$K$-ranked items as an important factor in improving the performance of top-$K$ recommendations, %take a different step %addressing the limitation of pairwise and listwise objectives. 

\subsection{Ranking Metrics for Top-$K$ Recommendations} \label{ssec:metrics}
%The performance of trained models should be validated with respect to appropriate metrics before being deployed. 
Here, we introduce four common evaluation metrics for top-$K$ recommendations. To explain ranking metrics, we represent the list of items sorted by the predicted scores for user $u$ as $\pi_u$, and the item at rank $k$ as $\pi_u(k)$. 
In addition, we define the $\hit(u, k)$ function that specifies whether $\pi_u(k)$ is in the validation dataset $\iset_u$. Explicitly,

\begin{equation} \label{eq:defhit}
    \hit(u, k) = \indicator[\pi_u(k) \in \iset_u]
\end{equation}
where $\indicator[\textit{statement}]$ is the indicator function, yielding $1$ if the \textit{statement} is true and $0$ otherwise. 

$K$-Truncated Precision (\precision{$K$}) and Recall (\recall{$K$}) are two of the most widely used evaluation metrics for top-$K$ recommendations. \precision{$K$} specifies the fraction of hit items in $\iset_u$ among recommended items. \recall{$K$} specifies the fraction of recommended items among the items in $\iset_u$. Notice that both metrics do not take into account the difference in the ranking of recommended items, while they collectively emphasize top-$K$-ranked items by counting only the items up to $K$-th ranks.

\begin{equation*}
    \begin{array}{l}
    \precision{K}(u, \pi_u) = \frac{1}{K} \sum_{k=1}^K \hit(u, k) \\ \\
    \recall{K}(u, \pi_u) = \frac{1}{|\iset_u|} \sum_{k=1}^K \hit(u, k)
    \end{array}
\end{equation*}

On the other hand, $K$-Truncated Normalized Discounted Cumulative Gain (NDCG)~\cite{ndcg-definition} and Truncated Average Precision (AP)~\cite{map-definition} take into account the relative ranks of items by weighting the impact of $\hit(u, k)$. Normalized DCG (NDCG) specifies a normalized value of $\text{DCG@}K$, which is divided by the ideal DCG $\text{IDCG@}K = \max_{\pi_u} \text{DCG@}K(u, \pi_u)$.

\begin{equation*}
    \begin{array}{l}
    \ndcg{K}(u, \pi_u)= \frac{\text{DCG@}K(u, \pi_u)}{\text{IDCG@}K} \\ \\
    \text{where } \ \text{DCG@}K(u, \pi_u) = \sum_{k=1}^{K} \frac{\hit(u, k)}{\log_2(k + 1)}.
    \end{array}
\end{equation*}

The truncated AP is defined as

\begin{equation*}
    \text{AP@}K(u, \pi_u) = \sum_{k=1}^{K}  
    \frac{\precision{k}(u, \pi_u) \hit(u, k)}{\min(K, |\iset_u|)}
\end{equation*}
where AP can be viewed as a weighted sum of $\hit$ for each rank $k=1,\ 2,\ \dots,\ K$, weighted by $\precision{k}$.  

We represent the aforementioned metrics in a unified way as $\obj(K)$ conditioned on weight functions $w(k, K)$,

\begin{equation} \label{eq:objective} 
\begin{aligned}
    \obj(K) &= \sum_{k=1}^K w(k, K) \hit(u, k) \\ 
            &= \begin{cases}
                       \precision{K} & \text{if}~ w(k, K)= K^{-1} \\
                       \recall{K} & \text{if}~ w(k, K) =  |\iset_u|^{-1} \\
                       \ndcg{K} & \text{if}~ w(k, K)= \frac{(\text{IDCG@}K)^{-1}}{\log (k+1)} \\ 
                       \text{AP@}{K} & \text{if}~ w(k, K) = \frac{\precision{k}}{\min(K, |\iset_u|)} \\
                  \end{cases}
\end{aligned}
\end{equation}
where we omit $(u, \pi_u)$ of the metrics for simplicity.

\section{Proposed Method} \label{sec:method}

In this section, we present two building blocks of our method, (1) explaining a factor based recommender with weighted hinge loss, and (2) introducing ranking metrics in terms of vector arithmetic as well as how to relax the metrics to be differentiable, which can be optimized using gradient descent. We call this relaxed metric DRM (Differentiable Ranking Metric). (3) Finally, we describe the learning procedure of the factor based recommender incorporated with DRM via joint learning.

%This section consists of two parts. First, we introduce matrix factorization with weighted hinge loss. Secondly, we represent ranking-based metrics with vector arithmetic and relax them to be differentiable. We conclude this section with the training procedure of \ourmodel{}.

\subsection{Factor Based Recommenders with Hinge Loss}
Factor based recommenders represent users and items in a latent vector space $\mathbb{R}^d$, and then formulate the preference score of user $u$ to item $i$ as a function of two vectors, user vector $\uv_u$ and item vector $\iv_i$. 
Dot product is one common method for mapping a pair of user and item vectors to a  predicted preference~\cite{BPR,PMF,WMF}. In~\cite{CML}, the collaborative metric learning (CML) embeds users and items in the euclidean space and defines its score function using the negative value of L2 distance of two factors. 

Our model uses either of dot product or L2 distance of user vector $\uv_u$ and item vector $\iv_i$ as a score function. 

\begin{equation} \label{eq:score}
    \hat{y}_{u,i} = \begin{cases}
                        \uv_u^T \iv_i & \text{Dot product}\\
                        -{\|\uv_u - \iv_i\|}^2 &\text{L2 distance}\\
              \end{cases}
\end{equation} 
Note that $\|\cdot\|$ is L2 norm. Having the score functions above, we update our model using weighted hinge loss in Eq.~\eqref{eq:hinge}. 
We calculate the weight $\Phi_{ui}$ of hinge loss by

\begin{equation} \label{eq:Phi}
    \Phi_{ui} = \log \left(1 + \frac{N}{|\mathcal{J}|} \sum_{j \in \mathcal{J}} \indicator[{1 - \hat{y}_{u, i} + \hat{y}_{u, j} \geq 0]}\right),
\end{equation}
similarly to negative sampling in \cite{CML}. 
For each update, 
$|\mathcal{J}|$ negative items are sampled from $\iset - \iset_u$ and used to estimate  $\Phi_{ui}$.

\subsection{Relaxed Precision} \label{ssec:neuralsort}
Sorting and ranking items can be seen as a permutation of items.  An $N$-dimension permutation corresponds to a vector ${p} = [p_1, p_2, \dots, p_N]^T$ where $p_i \in \{1, 2, \dots, N\}$ and $p_i \neq p_j$ if $i \neq j$.
For each vector $p$, we then have its permutation matrix $P \in\{ 0, 1 \}^{n \times n}$ and its element can be described as

\begin{equation} \label{eq:pmat}
    P_{i, j} = \begin{cases} 1 & \text{if}~ j = p_i \\0 & ~\text{otherwise.} \end{cases}
\end{equation}
For example, a permutation matrix $P = \begin{bmatrix}
\ 0 & 1 & 0\ \\
\ 1 & 0 & 0\ \\
\ 0 & 0 & 1\ \\
\end{bmatrix}$ maps a score vector ${v} = [3, 5, 1]^T$ to $P{v} =[5, 3, 1]^T$. 

In \cite{grover2018stochastic}, they propose continuous relaxation of sorting (namely \textit{NeuralSort}), which represents a sorting operation in a permutation matrix, and then relaxes the matrix into a continuous form. 
Specifically, sorting a vector $s=[s_1, s_2, \dots, s_N]^T$ in descending order can be represented in a permutation matrix such as

\begin{equation} \label{eq:perm}
    \perm{{s}}_{i, j} = \begin{cases} 1 & \text{if}~ j = \argmax [(n + 1 - 2i){s} - A_{{s}} \onecol]\\0 & ~\text{otherwise} \end{cases}
\end{equation} 
where $\onecol$ is the column vector having $1$ for all elements and $A_{{s}}$ is the matrix such that $A_{i, j} = |s_i - s_j|$. 
Then, $\softmax$ is used instead for relaxing the permutation matrix, i.e., 

\begin{equation} \label{eq:hperm}
    \hperm{{s}}_k = \softmax \left[ \tau^{-1}\left((n+1-2k){s} - A_{s} \onecol\right)\right]
\end{equation} 
where $\tau > 0$ is a temperature parameter. 
Larger $\tau$ values make each row of the relaxed matrix become flatter.
This transformation of \textit{NeuralSort} renders the permutation matrix in Eq.~\eqref{eq:perm} relaxed into a unimodal row-stochastic matrix, hence realizing the differentiation operation for sorting of real-value elements.
Eq.~\eqref{eq:hperm} is continuous everywhere and differentiable almost everywhere with respect to the elements of ${s}$. Furthermore, as $\tau \rightarrow 0^{+}$, $\hperm{{s}}$ reduces to the permutation matrix $\perm{{s}}$.

The $k$-th row $\perm{{s}}_k$ of the permutation matrix $\perm{{s}}$ is equal to the one-hot vector representation of the $k$-th ranked item. Thus, we can reformulate $\hit$ (Eq.~\eqref{eq:defhit}) using dot product of ${y}_u$ and $\perm{\hat{{y}}_u}_k$.

\begin{equation} \label{eq:neuhit}
    \hit(u, k) = {y}_u^T\perm{\hat{{y}}_u}_k
\end{equation}
Then, we obtain the representation of ranking metrics in  Eq.~\eqref{eq:objective} in terms of vector arithmetic using Eq.~\eqref{eq:neuhit}.

\begin{equation} \label{eq:vaobjective}
    \obj(K, w) = \sum_{k=1}^K w(k, K) {y}_u^T\perm{\hat{{y}}_u}_k
\end{equation}
By replacing $\perm{\hat{{y}}_u}$ in Eq.~\eqref{eq:vaobjective} with  differentiable $\hperm{\hat{{y}}_u}$, 
we obtain the differentiable relaxed objective, which can be used for optimization using gradient descent. 

\begin{equation} \label{eq:tildeobjective}
    \tilde{\obj}(K, w) = \sum_{k=1}^K w(k, K){{y}_u^T
    \hperm{\hat{{y}}_u}}
\end{equation}
%Since softmax is differentiable, this value is differentiable now.
%
%
%As such, nondifferentiable $P$ in the Eq.~\eqref{eq:vaobjective} has been relaxed to be differentiable $\tilde{P}$ in the Eq.~\eqref{eq:tildeobjective}, and therefore, we obtain that Eq.~\eqref{eq:tildeobjective} becomes differentiable.
We empirically find that the below equation is more stable in model training.

\begin{equation}\label{eq:neuloss}
   \neuloss = \| {y}_u - \sum_{k=1}^{K} w(k, K) \tilde{P_k}^{(\hat{{y}}_u)}\|^{2} 
\end{equation}
Note that minimizing Eq.~\eqref{eq:neuloss} is equivalent to maximizing Eq.~\eqref{eq:tildeobjective} since we have

\begin{equation*}
\begin{aligned}
    \tilde{\obj} &= {{y}_u}^T  \tilde{P}_{[1:K]} \\
    &= \frac{1}{2} \left[ {{y}_u}^T {y}_u + {\tilde{P}_{[1:K]}}^T \tilde{P}_{[1:K]} \right] - \frac{1}{2}\|{y}_u - \tilde{P}_{[1:K]}\|^2 \\ 
    &\geq -\frac{1}{2}\| {y}_u - \tilde{P}_{[1:K]}\|^2 = -\frac{1}{2}\neuloss  
\end{aligned}
\end{equation*} 
where $\tilde{P}_{[1:K]} = \sum_{k=1}^{K} w(k, K) \tilde{P_k}^{(\hat{{y}}_u)}$.

\subsection{Model Update} \label{ssec:training}
We incorporate the proposed objective in Eq.~\eqref{eq:neuloss} into the model learning structure (Eq.~\eqref{eq:hinge}) via joint learning of two objectives

\begin{equation} \label{eq:totobj}
    \begin{aligned}
    \ourloss &= \hingeloss + \lambda  \neuloss \\
             &= \sum_{u \in \uset} \sum_{i \in \iset_u} \sum_{j \in \iset - \iset_u} \Phi_{ui}[\mu - \hat{y}_{u, i} + \hat{y}_{u, j}]_{+} \\
         &+ \lambda \sum_{u \in \uset} \|{y}_u - \tilde{P}_{[1:K]}\|^2
    \end{aligned}
\end{equation}
%Eq.~\eqref{eq:totobj} is the objective of our model. 
where $\lambda$ is a scaling parameter to control the effect of $\neuloss$.

Similar to negative sampling~\cite{BPR,CML,NCF,LMF} used in factor based models using gradient descent updates, we follow a sampling procedure for positive items. We construct each training sample to contain a user $u$, a set of positive items $\mathcal{I}$ of size $\npos$ and a set of negative items $\mathcal{J}$ of size $\nneg$. Specifically, we construct a list of items ${y}_u$ of size $(\npos + \nneg)$ where first $\npos$ elements are from positive itemset $\mathcal{I}$, and next $\nneg$ elements are from negative itemset $\mathcal{J}$. Using~Eq.~\eqref{eq:score}, the list of predicted scores $\hat{{y}}_u$ is constructed similarly,

\begin{equation*}
    \begin{array}{l}
    {{y}}_u = {[{y}_{u, i_1}, {y}_{u, i_2}, \dots, {y}_{u, i_\npos}, {y}_{u, j_1}, \dots, {y}_{u, j_\nneg}]^T }  \\ \\
    \hat{{y}}_u = [\hat{y}_{u, i_1}, \hat{y}_{u, i_2}, \dots, \hat{y}_{u, i_\npos}, \hat{y}_{u, j_1}, \dots, \hat{y}_{u, j_\nneg}]^T.
    \end{array}
\end{equation*}

$\mathcal{J}$ is also used to calculate $\Phi_{ui}$ (Eq.~\eqref{eq:Phi}). The learning procedure for \ourmodel{} is summarized in Algorithm~\ref{algo:learning}. 

We adopt two regularization schemes. We first keep latent factors of users and items within the unit hypersphere, i.e., $\| \theta \| \leq 1$ where $\theta$ is either $\uv$ or $\iv$. 
For the model exploiting negative L2 distance as a score function, we also adopt covariance regularization~\cite{cogswell2015reducing} between all pairs of latent factors using a matrix $C_{i,j}= \frac{1}{|\Theta|} \sum_{\theta \in \Theta}(\theta_i - \mu_i)(\theta_j - \mu_j)^T$
where $\Theta = \{\uv_1, \uv_2, \dots, \uv_M, \iv_1, \iv_2, \dots, \iv_N\}$ and $\mu$ is the average vector of all user and item factors.
We define a regularization term $\loss_\text{C}$ as 

\begin{equation*}
    \loss_\text{C} = \frac{1}{|\Theta|}(\|C\|_f - \|\text{diag}(C)\|^2)
\end{equation*} where $\|\cdot\|_f$ is Frobenius norm. 
In case that L2 distance is used, we add this regularization to Eq.~\eqref{eq:totobj} with a control parameter $\lambda_\text{C}$, i.e., $\ourloss = \hingeloss + \lambda  \neuloss + \lambda_\text{C} \loss_{\text{C}}$.

\begin{center}
\DecMargin{1em} 
\begin{algorithm}[bt] 
\caption{Learning Procedure for \ourmodel{}}
\begin{algorithmic} \label{algo:learning}
\fontsize{10pt}{12pt}\selectfont
    \STATE Initialize user factors $\uv_u$ where $u \in \uset$
    \STATE Initialize item factors $\iv_i$ where $i \in \iset$

    \REPEAT
        \STATE Sample user $u$ from~$\uset$ \\
        \STATE Sample $\npos$ positive items $\mathcal{I}$ from $\iset_u$ \\ 
        \STATE Sample $\nneg$ negative items $\mathcal{J}$ from $\iset - \iset_u$ \\ 
        \STATE $\Delta \uv_u \leftarrow 0$
        \STATE $\Delta \iv_i \leftarrow 0$ for $i \in \mathcal{I}$  or $i \in \mathcal{J}$
        \STATE ${y}_u \leftarrow [{y}_{u, i_1}, \dots, {y}_{u, i_{\npos}}, {y}_{u, j_1}, \dots, {y}_{u, j_{\nneg}}]^T$ \\ 
        \STATE $\hat{{y}}_u \leftarrow [\hat{y}_{u, i_1}, \dots, \hat{y}_{u, i_{\npos}}, \hat{y}_{u, j_1}, \dots, \hat{y}_{u, j_{\nneg}}]^T$
        %\STATE $\mathbf{y}_u \leftarrow \{y_{u, i}| i \in \mathcal{I} \} \cup \{y_{u, j}| j \in \mathcal{J} \}$
        %\STATE $\hat{\mathbf{y}}_u \leftarrow \{\hat{y}_{u, i}| i \in \mathcal{I} \} \cup \{ \hat{y}_{u, j}| j \in \mathcal{J} \}$
        \STATE Choose one positive item $i$ with smallest $\hat{y}_{u, i}$ among $\mathcal{I}$
        \STATE Choose one negative item $j$ with largest $\hat{y}_{u, j}$ among $\mathcal{J}$
        \FOR{$\theta \leftarrow \{\uv_u, \iv_i, \iv_j \}$}{
            \STATE $\Delta\theta  \leftarrow \Delta\theta + \nabla \hingeloss$
        }\ENDFOR
        
        \FOR{$\theta \leftarrow \{\uv_u, \iv_{i_1}, \dots, \iv_{i_\npos}, \iv_{j_1}, \dots, \iv_{j_\nneg} \}$}{
            \STATE $\Delta\theta \leftarrow \Delta\theta + \lambda \nabla \neuloss$
        }\ENDFOR
        
        \FOR{$\theta \leftarrow \{\uv_u, \iv_{i_1}, \dots, \iv_{i_\npos}, \iv_{j_1}, \dots, \iv_{j_\nneg} \}$}{
            \STATE Update $\theta$ with $\Delta \theta$ using Adagrad Optimizer
            \STATE $\theta \leftarrow \theta / \min(1, \|\theta\|)$ 
        }\ENDFOR
    \UNTIL{Converged} 
\end{algorithmic} 
\label{alg:learn-mf}
\end{algorithm}
\IncMargin{1em} 
\end{center}

\section{Empirical Evaluation} \label{sec:eval}
In this section, we evaluate our proposed model in comparison with state-of-the-art recommendation models.

\subsection{Experiment Setup} \label{ssec:eval-protocol}
We test our approach and baseline models upon four real-world datasets. 
%\begin{itemize}
%    \item \textbf{SketchFab}~\cite{sketchfab} contains users' \textit{likes} on 3D models. We treat each \textit{like} as a user-item interaction. 
%    \item \textbf{Epinion}~\cite{epinion1} contains 5-star rating reviews of customers on products from a web service. We view each rating as a user-item interaction.
%    \item \textbf{ML-20M}~\cite{movielens} contains the movie ratings from a move recommendation service. Ratings are from 0.5 to 5.0 in 0.5 increments. We treat ratings as positive interactions and exclude the ratings lower than 4. 
%    \item \textbf{Melon}\footnote{https://arena.kakao.com/c/7} contains playlists from a music streaming service. To be consistent with the implicit feedback setting, we treat each playlist as a user, and songs in a playlist as a list of positive items of the user. 
%\end{itemize}
\textbf{SketchFab}~\cite{sketchfab} contains users' \textit{likes} on 3D models. We treat each \textit{like} as a user-item interaction. 
\textbf{Epinion}~\cite{epinion1} contains 5-star rating reviews of customers on products from a web service. We view each rating as a user-item interaction. 
\textbf{ML-20M}~\cite{movielens} contains the movie ratings from a movie recommendation service. Ratings are from 0.5 to 5.0 in 0.5 increments. We treat ratings as positive interactions and exclude the ratings lower than 4. 
\textbf{Melon}\footnote{https://arena.kakao.com/c/7} contains playlists from a music streaming service. To be consistent with the implicit feedback setting, we treat each playlist as a user, and songs in a playlist as a list of positive items of the user. 
The dataset statistics are summarized in Table~\ref{tbl:data-stats}.

\begin{table}[hb]
\centering
\begin{adjustbox}{width=\linewidth}
\begin{tabular}{lcccc}
    \hline
                    & SketchFab    & Epinion        & ML-20M       & Melon \\ \hline
    \#users        & 15.5K        & 21.3K          & 136K      & 104K  \\
    \#items        & 28.8K        & 59.2K          & 15.5K      & 81.2K      \\
    \#interactions & 558K       & 631K         & 9.98M     & 4.10M \\
    Avg. row        & 35.9        & 29.4           & 73.0       & 39.2     \\
    Avg. col        & 19.4        & 10.6           & 643   & 50.5  \\
    Density         & 0.12\%       & 0.048\%         & 0.47\%     & 0.048\%  \\
    \hline
\end{tabular} 
\end{adjustbox}
\caption{
Dataset statistics. \#users, \#items, and \#interactions denote the number of users, items, and interactions, respectively;
avg. row and avg. col denote the average number of items that each user has interacted with, and the average number of users who have interacted with each item respectively;
density denotes the interaction matrix density (i.e., density = \#interactions / (\#users $\times$ \#items);
}
\label{tbl:data-stats}
\end{table}

\paragraph{Evaluation protocol} \label{para:eval}
We split the interaction dataset into training, validation, and test datasets in 70\%, 10\%, and 20\% portions, respectively. 
For each model, we first train it once using its training dataset to find the best hyperparameter settings, while evaluating the hyperparameter settings using its validation dataset. 
We then train the model 5 times with the best hyperparameter settings using the training and validation datasets and evaluate the model using its test dataset, thereby reporting the average performance in evaluation metrics. 
We use \recall{50} for model validation, and  we report mean AP@10 (\map{10}), \ndcg{10}, \recall{50}, and \ndcg{50}. 
Each metric is averaged for all users except for users with fewer than five interactions in the training dataset, and users with no interactions in the test dataset.
We conduct Welch's T-test~\cite{ttest} on results and denote the best results with $p$-value lower than $0.01$ in boldface.

%, with considerations of the recent trend of two-stage recommenders~\cite{netflixrecommender,youtuberec}. Recommenders can be used in (1) the candidate generation stage. At this stage, we are interested in filtering items that are less likely to be recommended. Filtered items are further ranked in the next stage at (2) the ranking stage. At this stage, we rank items according to preferences generated by ranker models. Recall metric is better used to evaluate the performance of recommenders in candidate generation since the ranking is not considered in this stage. For the ranking stage, we exploit ranking based evaluation metrics, such as MAP and NDCG, with a small truncation value $10$ because only a handful of items is recommended to users. For the candidate generation stage, we use large truncation $50$.

\paragraph{Our models} \label{ssec:ours}
We use two variants of \ourmodel{} with different score functions. \ourmodel{dot} uses dot product, while \ourmodel{L2} exploits the negative value of L2 distance as a score function. We set $w(k, K)$ to be $1$ for ease of computation.

\paragraph{Baseline models} \label{ssec:eval-models}
We compare \ourmodel{} with the following baselines. Note that except for SLIM, all baselines fall in the category of factor based recommenders.
\begin{itemize}
    \item \textbf{SLIM}~\cite{ning2011slim} is a state-of-the-art item-item collaborative filtering algorithm in which the item-item similarity matrix is represented as a sparse matrix. It poses L1 and L2 regularization on the item-item similarity matrix. %The regularization is controlled $\lambda_1$, $\lambda_2$, respectively.
    
    \item \textbf{CDAE}~\cite{DAE} can be seen as a factor based recommender where user factors are generated via an encoder. The encoder takes embeddings of items the user has consumed and the embedding of the user as input and returns the latent factor of the user. We use a neural network with no hidden layers as described in the original paper implementation.
    
    \item \textbf{WMF}~\cite{WMF} uses mean squared error as the objective function and minimizes it using alternating least squares. We put the same L2 regularization on both user and item factors.
    
    \item \textbf{BPR} \cite{BPR} exploits the pairwise sigmoid objective. We put the same L2 regularization on both user and item factors.
    
    \item \textbf{WARP}~\cite{warp,weston2013learning} is trained using hinge loss with approximated rank based weights.
    
    \item \textbf{CML}~\cite{CML} models user-item preference as a negative value of the distance between the user vector and the item vector. It poses regularization on the latent factors that are distributed with the same density in the unit hypersphere using Frobenius norm of the covariance of the latent factors~\cite{cogswell2015reducing}.
    
    \item \textbf{SQLRank}~\cite{SQLRANK} views a recommendation problem as sorting lists. Then it optimizes the upper bound of log-likelihood of probabilities of correctly sorted lists.

    \item \textbf{SRRMF}~\cite{SRRMF} is an extension of WMF. It interpolates  unobserved feedback scores to be nonzero so as to treat each negative item differently.
    
\end{itemize}
We use an open-source implementation Implicit~\cite{implicit} for WMF and BPR, and another open-source implementation LightFM~\cite{lightfm} for WARP. We use implementations publicly available by the authors for SLIM\footnote{https://github.com/KarypisLab/SLIM}, SQLRank\footnote{\url{https://github.com/wuliwei9278/SQL-Rank/}} and SRRMF\footnote{\url{https://github.com/HERECJ/recsys/tree/master/}}. We implement CDAE, CML, and \ourmodel{} using Python 3.7.3 and PyTorch 1.5.0. We run our experiment on a machine with an Intel(R) Xeon(R) CPU E5-2698, 160G memory, and an NVIDIA Tesla V100 GPU with CUDA 10.1.

Our implementation of DRM is available and on GitHub\footnote{\url{https://github.com/ita9naiwa/DRM-recsys}}.

\subsubsection{Hyperparameter settings} \label{ssec:hyperparams}
\begin{table}[ht]
\begin{adjustbox}{width=\linewidth}
\begin{tabular}{c|c|c}
\hline
\multirow{2}{*}{Model} & \multirow{2}{*}{Hyperparameter} & \multirow{2}{*}{Search space} \\ & & \\ \hline 
\multirow{2}{*}{SLIM} & 
L1 regularization & 
\multirow{2}{*}{$\{ 0.1, 0.5, 1.0, 3.0, 5.0, 10.0, 20.0 \}$} \\ \cline{2-2} & 
 L2 regularization &  \\ \hline
\begin{tabular}[c]{@{}c@{}}all models \\ except for SLIM\end{tabular} & latent factor dim. $d$ & $\{ 16, 32, 64, 128 \}$ \\ \hline
\begin{tabular}[c]{@{}c@{}}CDAE, BPR, \\ WMF, SRRMF\end{tabular} & L2 regularization & $\{ 1\text{e-}4, 1\text{e-}3, 3\text{e-}3, 0.01, 0.03 \}$ \\ \hline
\begin{tabular}[c]{@{}c@{}}CDAE, BPR, WARP, \\ CML, SQLRANK, \\ \ourmodel{dot}, \ \ourmodel{L2}\end{tabular} & learning rate& $\{ 1\text{e-}4, 1\text{e-}3, 5\text{e-}3, 0.01, 0.03, 0.05, 0.1 \}$ \\ \hline
WMF, SRRMF & confidence weight $c$& $\{$1.0, 3.0, 5.0, 10.0, 20.0$\}$ \\ \hline
WARP, CML & \begin{tabular}[c]{@{}c@{}}negative sampling \\ $|\mathcal{J}|$ \end{tabular} & $\{$10, 20, 30, 40, 50$\}$ \\ \hline
\begin{tabular}[c]{@{}c@{}}WARP, CML,  \\ \ourmodel{dot}, \ \ourmodel{L2} \end{tabular} & margin $\mu$ & $\{1\}$ \\ \hline
\multirow{3}{*}{\ourmodel{dot}, \ \ourmodel{L2}} &temperature $\tau$ & $\{ 0.1, 0.3, 0.5, 1.0, 3.0, 10.0 \}$ \\ \cline{2-3}
 & positive sampling $\npos$ & $\{$1, 2, 3, 4, 5$\}$ \\ \cline{2-3}
 & scaling parameter $\lambda$ & $\{$0.1, 0.5, 1.0, 2.0, 3.0, 5.0$\}$ \\ \hline
 CML, \ourmodel{L2} & covariance regularization $\lambda_c$& $\{$0.1, 0.5, 1.0, 3.0, 5.0, 10.0$\}$ \\ \hline

\end{tabular}
\end{adjustbox}
\caption{Hyperparameter search space. }
\label{tbl:hyperparam}
\end{table}

We tune hyperparameters using grid-search. The hyperparameter search spaces are described in Table \ref{tbl:hyperparam}. We use Adagrad Optimizer for CDAE and CML, and use stochastic gradient descent for BPR as mentioned in the original papers. We use Adagrad optimizer for WARP, \ourmodel{dot}, and \ourmodel{L2}. The number of negative sampling $\nneg$ for \ourmodel{} variants is set to be $15 \cdot \npos$.

\begin{table*}[ht]
    \begin{adjustbox}{width=\textwidth}
        \begin{tabular}{cc|cccccccc|cccc}
\hline
\multirow{2}{*}{Datasets} & \multirow{2}{*}{Metrics} & \multicolumn{8}{c}{Baselines} & \multicolumn{2}{|c}{Our models} \\ 
     &  & SLIM & CDAE & BPR & WMF & WARP & CML & SQLRANK & SRRMF & \ourmodel{dot} & \ourmodel{L2} \\ \hline
\multirow{4}{*}{SketchFab} & $\map{10}$ & 0.0300 & 0.0351 & 0.0216 & 0.0335 & 0.0363 & 0.0358 & 0.0101 & 0.0200 & \textbf{0.0399} (9.9\%) & \textbf{0.0390} (7.4\%)\\
 & $\ndcg{10}$ & 0.1163 & 0.1301 & 0.0905 & 0.1257 & 0.1354 & 0.1379 & 0.0417 & 0.0862 & \textbf{0.1479} (7.2\%) & \textbf{0.1466} (6.3\%)\\
 & $\recall{50}$ & 0.2696 & 0.2793 & 0.2168 & 0.2862 & 0.2923 & {0.3040} & 0.1422 & 0.1550 &\textbf{0.3091} (1.6\%) & {0.3028} (-0.3\%)\\ 
 & $\ndcg{50}$ & 0.1067 & 0.1657 & 0.1218 & 0.1645 & 0.1724 & 0.1778 & 0.0537 & 0.0995 &\textbf{0.1855} (4.3\%) & \textbf{0.1836} (3.2\%)\\ \hline
\multirow{4}{*}{Epinion} & $\map{10}$ & 0.0086 & 0.0128 & 0.0062 & 0.0123 & 0.0100 & 0.0130 & 0.0036 & 0.0107 & \textbf{0.0144} (10.7\%) & \textbf{0.0137} (5.3\%)\\
 & $\ndcg{10}$ & 0.0357 & 0.0453 & 0.0238 & 0.0486 & 0.0387 & 0.0493 & 0.0168 & 0.0428 & \textbf{0.0532} (7.9\%) & \textbf{0.0523} (6.0\%)\\
 & $\recall{50}$ & 0.1081 & 0.1123 & 0.0661 & 0.1325 & 0.1158 & {0.1347} & 0.0432 & 0.1275 & {0.1361} (1.0\%) & 0.1308 (-2.8\%)\\ 
 & $\ndcg{50}$ & 0.0410 & 0.0646 & 0.0362 & 0.0726 & 0.0610 & 0.0736 & 0.0252 & 0.0680 & \textbf{0.0766} (4.0\%) & {0.0746} (1.3\%)\\ \hline
 \multirow{4}{*}{ML-20M} & $\map{10}$ & 0.1287 & 0.1569 & 0.0787 & 0.1034 & 0.1284 & 0.1331 & \multirow{4}{*}{-} & 0.0987 & 0.1475 (-5.9\%)& \textbf{0.1598} (1.8\%) \\
 & $\ndcg{10}$ & 0.2761 & 0.3205 & 0.1917 & 0.2561 & 0.2750 & 0.2824 &  & 0.2532 & 0.3068(-4.2\%) & \textbf{0.3267} (1.9\%) \\
 & $\recall{50}$ & {0.4874} & {0.4829} & 0.3431 & 0.4676 & 0.4791 & {0.4874} &  & 0.4786 &  \textbf{0.4944} (1.4\%)& \textbf{0.5014} (2.8\%) \\
 & $\ndcg{50}$   & 0.2511 & 0.3667 & 0.2394 & 0.3288 & 0.3360 & 0.3416 &  & 0.3244 & 0.3627 (-1.0\%) & \textbf{0.3912} (6.6\%) \\\hline
\multirow{4}{*}{Melon} & $\map{10}$ & 0.0838 & 0.0612 & 0.0400 & 0.0562 & 0.0474 & 0.0692 & \multirow{4}{*}{-} & 0.0652 & 0.0764 (-8.8\%)& \textbf{0.0892} (6.4\%) \\
 & $\ndcg{10}$ & 0.1768 & 0.1041 & 0.0972 & 0.1303 & 0.1217 & 0.1659 &  & 0.1324 & \textbf{0.1802} (1.9\%) & \textbf{0.2010} (13.6\%) \\
 & $\recall{50}$ & 0.3415 & 0.1928 & 0.2159 & 0.2537 & 0.2577 & 0.3361 &  & 0.2863 & \textbf{0.3471} (1.6\%) & \textbf{0.3700} (8.3\%) \\
 & $\ndcg{50}$ & 0.2206 & 0.1335 & 0.1363 & 0.1716 & 0.1654 & 0.2206 &  & 0.1842 & \textbf{0.2334} (5.8\%)& \textbf{0.2550} (15.5\%) \\\hline
        \end{tabular} 
    \end{adjustbox}
    \caption{Recommendation performance of different models. The best results with $p \leq 0.01$ using paired T test are boldfaced.}
    \label{tbl:overall}
\end{table*}

\subsection{Quantitative Results} \label{ssec:quantitativeresults}

Table~\ref{tbl:overall} shows the recommendation performance of the baseline and \ourmodel{} variant models on different datasets. 

When comparing models with the same structure, which use the same score function or share the objectives, we observe that \ourmodel{} consistently outperforms the respective models  with the same structure. 
For example, WARP and \ourmodel{dot} use the same score function and share hinge loss, but \ourmodel{dot} consistently outperforms WARP for all the datasets. This pattern recurs for CML and \ourmodel{L2}, whose score function is L2 distance and share hinge loss. This clarifies that the proposed \ourmodel{} objective leads factor based models to make better recommendations by exploiting the top-$K$ recommendation nature. 

We observe \ourmodel{L2} significantly improves performance over the baselines by up to 15.5\% on SketchFab, Epinion, and Melon datasets in most cases  except for \recall{50} on SketchFab and Epinion. Also, we observe that the performance gain is significant in \ndcg{10} and \map{10}, except for ML-20M. We interpret it as \ourmodel{} tends to perform better than the baselines when the dataset is sparse, i.e., the number of interactions is insufficient to learn useful latent factors of users and items.

We record the performance of SQLRank only upon SketchFab and Epinion datasets, since it is hardly possible for us to train SQLRank successfully with large datasets, ML-20M and Melon, due to the huge training time. 
One interesting thing to note is that among the models using hinge loss, the models using dot product as a score function (WARP and \ourmodel{dot}) show better performance than the models using negative L2 distance  (CML and \ourmodel{dot}) on SketchFab and Epinion datasets, which are relatively small datasets. In contrast, the models exploiting negative L2 distance as a score function outperform the models using dot product on larger datasets, ML-20M and Melon.
 
\subsection{Exploratory Analysis}
\paragraph{Effects of positive sampling} \label{para:possample}
\begin{figure}[ht] 
    \centering
    \begin{subfigure}[ht]{0.99\linewidth}
    \includegraphics[width=\linewidth]{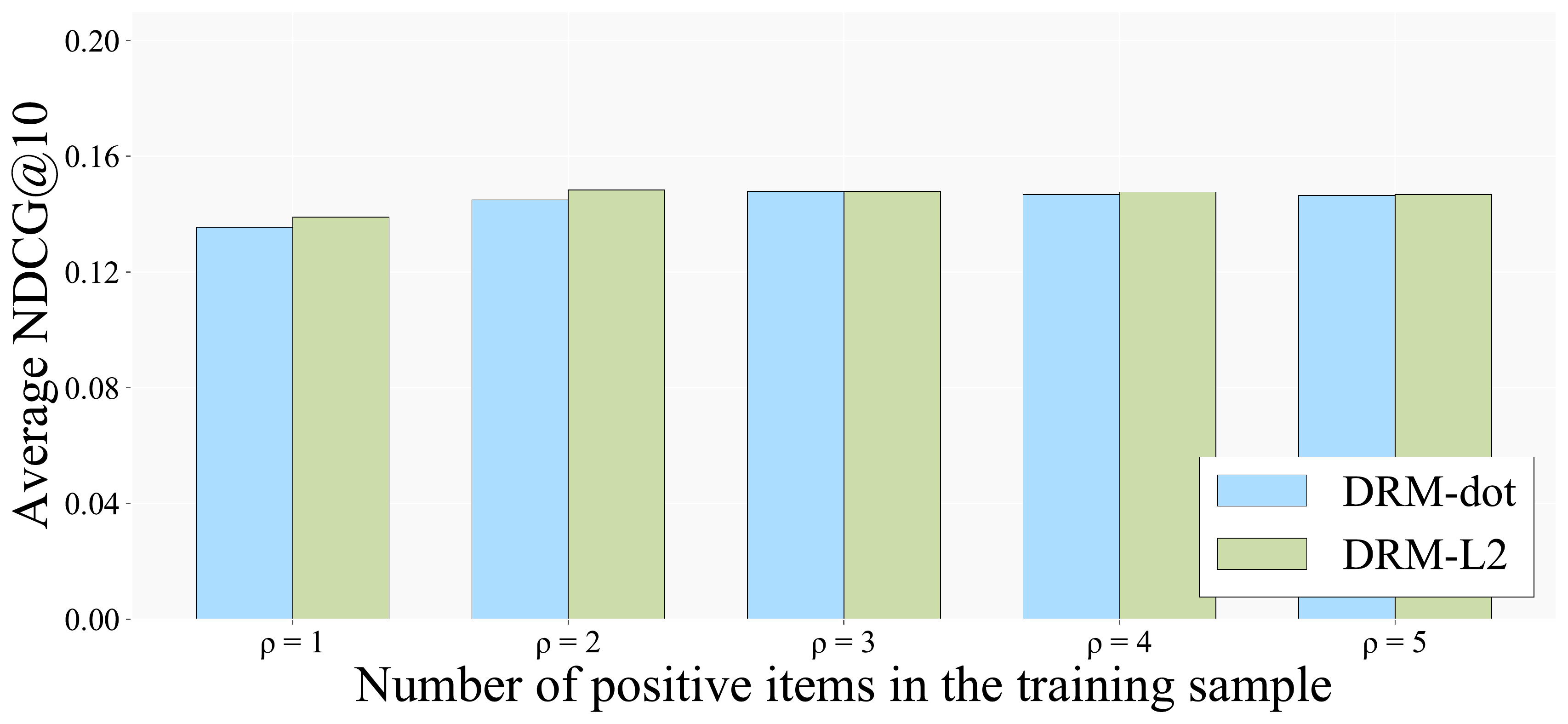}
    \caption{SketchFab}
    \end{subfigure}\\
    \begin{subfigure}[ht]{0.99\linewidth}
    \includegraphics[width=\linewidth]{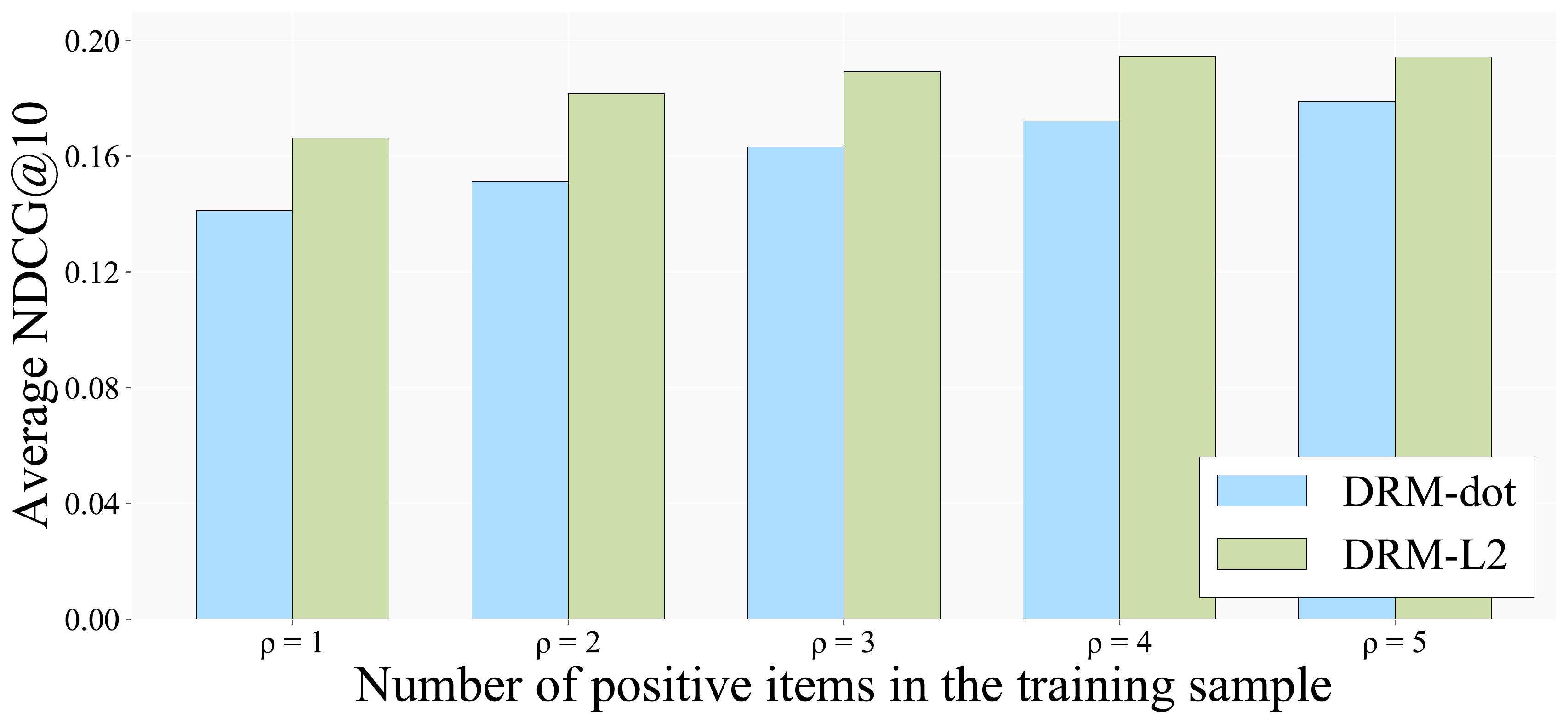}
    \caption{Melon}
    \end{subfigure}
    \caption{Effects of the number of positive samples $\npos$. Scaling parameter $\lambda$ is set to be $1$.}

    \label{fig:hypereffects}
\end{figure}
The training procedure of \ourmodel{} requires sampling $\npos$ positive items. In Fig.~\ref{fig:hypereffects}, we experiment on SketchFab and Melon datasets to see the relation between the number of positive samples $\npos$ and the recommendation performance.  As the number of positive items $\npos$ increases, \ndcg{10} improves but this positive effect often decreases gradually for large $\npos$ beyond some point, e.g., $\npos = 3$ for SketchFab.

\paragraph{Performance over different user groups}
\begin{figure}[ht] 
    \centering
    \begin{subfigure}[ht]{0.99\linewidth}
    \includegraphics[width=\linewidth]{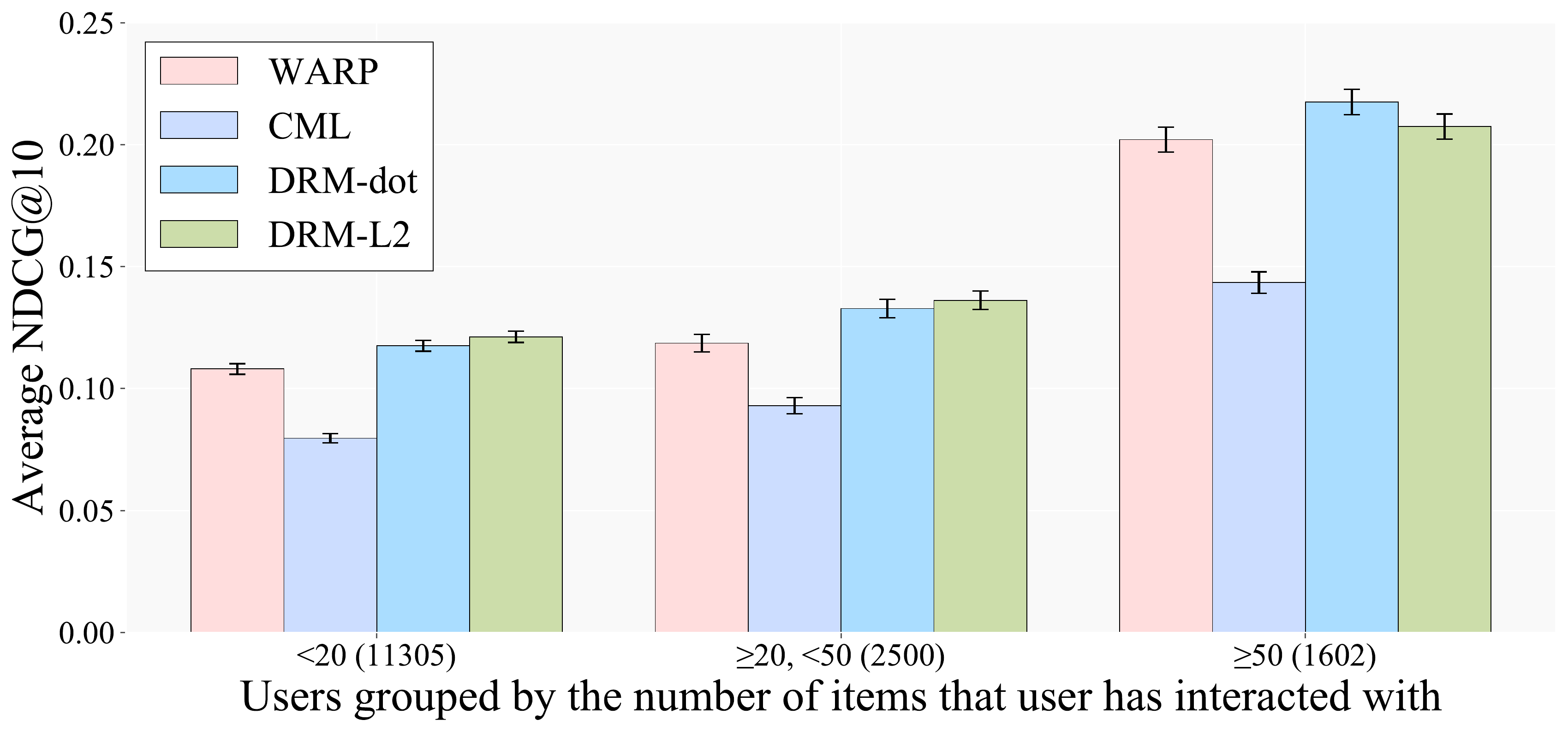}
    \caption{SketchFab}
    \end{subfigure}\\
    \begin{subfigure}[ht]{0.99\linewidth}
    \includegraphics[width=\linewidth]{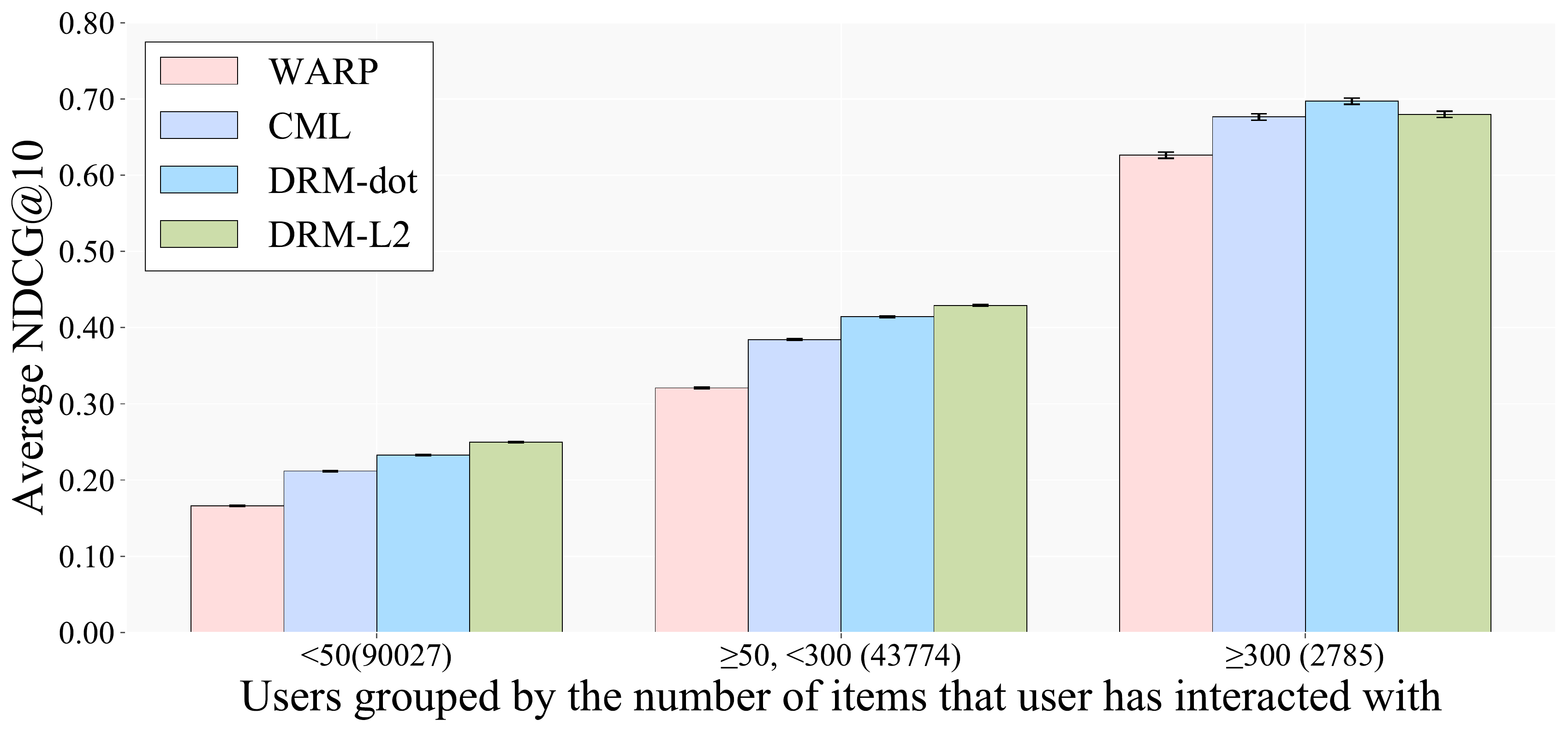}
    \caption{ML-20M}
    \end{subfigure}
    \caption{\ndcg{10} among different user groups by the number of interactions. The numbers in the parenthesis denote the number of users in each group.}

    \label{fig:groupndcg}
\end{figure}

Fig.~\ref{fig:groupndcg} shows average $\ndcg{10}$ of users grouped by the number of interactions in the training datasets. \ourmodel{} objective consistently improves recommendation performance for all user groups on both SketchFab and ML-20M, especially when the number of interactions of users is relatively small. This indicates the robustness of our model against sparse datasets. 

\begin{table}[ht] 
\centering
\begin{tabular}{cccl} 
\hline
 & CML & DRM-only & \ourmodel{L2} \\ \hline
MAP@10 & 0.0358 & 0.0310 & \textbf{0.0390} \\
NDCG@10 & 0.1379 & 0.1234 & \textbf{0.1466} \\
Recall@50 & {0.3040} & 0.2802 & {0.3028} \\
NDCG@50 & 0.1778 & 0.1608 & \textbf{0.1836} \\ \hline
\end{tabular} 
\caption{Performance comparison of three models with the same score function but different objectives on SketchFab.}
\label{tbl:drmonly}
\end{table}

\paragraph{Effects of DRM objective} \label{para:drm}
\begin{table}[ht]
\begin{adjustbox}{width=\linewidth}
\begin{tabular}{ccccc}
\hline
ML-20M & \multicolumn{2}{c}{Training Dataset} & \multicolumn{2}{c}{Validation Dataset} \\
Metrics & MAP@10 & NDCG@10 & MAP@10 & NDCG@10 \\ \hline
MSE & -0.957 & -0.946 & -0.943 & -0.745 \\
WARP & -0.983 & -0.983 & -0.978 & -0.915 \\ \hline
MSE + DRM & -0.984 & -0.972 & -0.976 & -0.946 \\
\ourmodel{dot} & -0.985 & -0.984 & -0.936 & -0.970 \\ \hline
\end{tabular}
\end{adjustbox}
\caption{Correlation coefficients between training loss and \map{10}/\ndcg{10} during 20 epochs except for first two epochs where the variation is huge. MSE denotes a factor based recommender with the mean squared error objective.}
\label{tbl:alignment}
\end{table}
In Table~\ref{tbl:drmonly}, we compare three models, CML, \ourmodel{L2}, and the model trained using DRM objective only (\ourmodel{}-only), which exploit the same score function. 
\ourmodel{L2} shows the best performance, and \ourmodel{}-only performs worse than the other two models. This result clarifies the benefit of joint learning of hinge loss and \ourmodel{} objective. 
We conjecture that model training using only a listwise objective might suffer from a lack of training data because only one sample exists for a user. \ourmodel{} overcomes this insufficiency using the joint learning with hinge loss, enjoying benefits from both pairwise and listwise objectives.

We also conduct an illustrational experiment to answer the question whether the DRM objective is more closely related to the ranking metrics than other objectives. Table~\ref{tbl:alignment} describes the correlation coefficients between training loss and \map{10} or \ndcg{10} on ML-20M training and validation datasets. 
MSE denotes a factor based recommender with the mean squared error objective. 
MSE + DRM denotes the model trained using the sum of mean squared error and \ourmodel{} as an objective. 
As we claimed, loss values are more closely related to DRM objective than hinge loss or mean square error. In other words, optimizing \ourmodel{} objective directly increases model performance. This correlation gap is significant on validation datasets, e.g., $-0.745$ (MSE) versus $-0.946$ (MSE + DRM) of \ndcg{10}.

\section{Related Work} \label{sec:relatedwork}
The direct optimization of ranking metrics is considered as an important problem in the domain of information retrieval~\cite{ltrsurvey}. One of the most common approaches for the problem is to optimize nondifferentiable ranking metrics using differentiable surrogates, such as the upper bound of the ranking metrics. In~\cite{approxrank}, authors derive relaxation of $\hit$ for AP and NDCG with a unified view of ranking metrics and propose an optimization of the relaxed ranking metrics. SoftRank~\cite{taylor2008softrank} defines expected NDCG over the distribution of scores and optimizes the expected NDCG. 
However, they are not tailored for truncated ranking metrics and do not fit well for top-$K$ recommendations. Different from these methods, \ourmodel{} relaxes truncated ranking metrics directly, in a more appropriate way for top-$K$ recommendations.

Listwise Collaborative Filtering~\cite{listwisecollaborativefiltering} addresses the misalignment issue between cost and objective on $K$-Nearest neighbors recommenders. Specifically, this work exploits the similarity between two lists for $K$-Nearest neighbors recommenders. Our work is complementary since we support factor based models, which is not memory based. 
SQLRANK~\cite{SQLRANK} is a factor based recommender minimizing the upper bound of negative log-likelihood of the permutation probability. 
In spite of its theoretical soundness, we are not able to achieve sufficient performance in our experiments with SQLRANK. The similar pattern to our experiments is also shown in \cite{SRRMF}. 
It is conjectured that \ourmodel{} achieves competitive performance by making use of both pairwise and listwise objectives, differently from SQLRANK that focuses on the theoretical foundation of listwise objective based model training.

\section{Conclusion}
In this work, we present \ourmodel{} a differentiable ranking metric, by which we directly optimize ranking metrics and hence improve the performance of top-$K$ recommendations. 
We show that the \ourmodel{} objective is readily incorporated into the existing factor based recommenders via joint learning. 
Through experiments, we show that DRM considerably outperforms other recommenders on  real-world datasets. Also, we demonstrate that optimizing our objective is more closely correlated to maximizing recommendation performance.

%Our future work is to make the theoretical analysis of \ourmodel{} and applying \ourmodel{} to other ranking metrics than the Precision we explore in this work. It is also interesting to apply the DRM to deep learning recommenders such as AutoEncoders.
 
\ifx\tosubmitashidden\undefined

\else
\fi 
\bibliography{main}
\end{document}

\maketitle

\section{Implementation Details}

\section{Math Details}
%%% 식 고치려고 잠깐 state도 가져옴
$\textbf{Interpretation.}$ We have the gradient update rule $\nabla \neuloss$ of the loss \eqref{eq:neuloss} with respect to a latent vector $\mathbf{x}$ such that $\mathbf{x}=\alpha$ or $\mathbf{x} = \beta_j$ for all $j$:

\begin{equation} \label{eq:clr-update}
\begin{aligned}
\nabla\neuloss = \frac{2}{\tau} \left(\frac{\partial {\hat{y_u}}}{\partial \mathbf{x}} \right)^T \sum_{i} W^{(i)} \left(\tilde{P}_{[1:K]} - y_u \right)
\end{aligned}
\end{equation}
where  
\begin{equation*}
\text{sgn}_{ij} =\begin{cases} 1 & \text{if } \hat{y}_{u, i}>\hat{y}_{u, j} \\ 0 & \text{if } \hat{y}_{u, i}=\hat{y}_{u, j}\\-1 & \text{if } \hat{y}_{u, i}<\hat{y}_{u, j}\end{cases}, \\
\end{equation*}
and\\
\begin{equation*}
\begin{aligned}
z^{(i)} &= \tau^{-1} \left((n+1-2i)\hat{y_u}-A_{\hat{y_u}} \mathbf{1} \right),\\ 
W^{(i)} &= (H^{(i)} (D^{(i)} + R))^T, \\
H^{(i)} &= \text{diag}(\text{softmax}(z^{(i)})) - \text{softmax}(z^{(i)})(\text{softmax}(z^{(i)}))^T , \\
D^{(i)} &= \text{diag}((n+1-2i)\mathbf{1}),\\ 
R &= \text{diag} \left(\left(-\sum_{k; k \neq j}\text{sgn}_{jk}\right)\mathbf{1} \right) + \begin{bmatrix} \text{sgn}_{jk} \end{bmatrix}_{n \times n}\\
\end{aligned}
\end{equation*}
for all $j$. \\
%%%

$\textit{Sketch of the derivation.}$  \\
\noindent $(i)$ Let $Q := \tilde{P}_{[1:K]}$ with $\tilde{P_i} = \begin{bmatrix} \tilde{P_{i1}} \ \tilde{P_{i2}} \ \cdots \ \tilde{P_{in}}\end{bmatrix}^T$. 
For convenience, let $s=\hat{y_u}$ with $s_i = \hat{y_{u,i}}$. By Chain Rule, 
$$
\frac{\partial \neuloss}{\partial \mathbf{x}} = \frac{\partial \neuloss}{\partial Q} \left( \sum_{i} \frac{\partial Q}{\partial \tilde{P_i}} \frac{\partial \tilde{P_i}}{\partial s}\right)\frac{\partial s}{\partial \mathbf{x}}
.$$
Since $\mathcal{L}_\text{neu} = (y_u - Q)^2$, we can deduce 
$$
\frac{\partial \mathcal{L}_\text{neu}}{\partial Q}=2(Q-y_u)^T .
$$\\
\noindent $(ii)$ By the definition of Q, we can derive 
\begin{equation*}
\begin{aligned}
\frac{\partial Q}{\partial \tilde{P_i}} &= \begin{bmatrix} \frac {\partial(\sum_j \tilde{P_{j1}})}{\partial \tilde{P_{i1}}} & \frac {\partial(\sum_j \tilde{P_{j1}})}{\partial \tilde{P_{i2}}} & \cdots & \frac {\partial(\sum_j \tilde{P_{j1}})}{\partial \tilde{P_{in}}} \\ \frac {\partial(\sum_j \tilde{P_{j2}})}{\partial \tilde{P_{i1}}} & \frac {\partial(\sum_j \tilde{P_{j2}})}{\partial \tilde{P_{i2}}} & \cdots & \frac {\partial(\sum_j \tilde{P_{j2}})}{\partial \tilde{P_{in}}} \\
\vdots & \vdots & \ddots & \vdots \\ \frac 
{\partial(\sum_j \tilde{P_{jn}})}{\partial \tilde{P_{i1}}} & \frac {\partial(\sum_j \tilde{P_{jn}})}{\partial \tilde{P_{i2}}} & \cdots & \frac {\partial(\sum_j \tilde{P_{jn}})}{\partial \tilde{P_{in}}} \end{bmatrix} \\ &= \begin{bmatrix} 1 & 0 & \cdots & 0 \\ 0 & 1 & \cdots & 0\\
\vdots & \vdots & \ddots &\vdots \\0 & 0 & \cdots & 1\end{bmatrix}_{n \times n}%; \text{ the identity matrix}
\end{aligned}
\end{equation*}
for all $i=1,\ 2,\ 3,\ \cdots,\ n$.\\

\noindent $(iii)$ Set $z^{(i)}:=\tau^{-1}((n+1-2i)s - A_s \mathbf{1}) \in \mathbb{R}^{n}$. With the property $|s_i - s_j| = \text{sgn}_{ij}(s_i-s_j)$, we can write $A_s \mathbf{1}$ as
%$$
%A_s \mathbf{1} = \begin{bmatrix} \sum_{k;k \neq 1} \text{sgn}_{1k}(s_1-s_k)\\ \sum_{k;k \neq 2} \text{sgn}_{2k}(s_2-s_k) \\ \vdots \\ \sum_{k;k \neq n} \text{sgn}_{nk}(s_n-s_k)\end{bmatrix}_{n \times 1},
%$$
%from the definition of $A_s$. Thus
\begin{equation*}
(A_s \mathbf{1})_j=\sum_{k;k \neq j} \text{sgn}_{jk}(s_j-s_k). 
\end{equation*}
Hence the $j$-th component ${z^{(i)}}_j$ of $z^{(i)}$ is 
\begin{equation*}
{z^{(i)}}_j=(n+1-2i)s_j - \sum_{k;k \neq j} \text{sgn}_{jk}(s_j-s_k).
\end{equation*}\\
$(iv)$ Note that $\frac{\partial \tilde{P_i}}{\partial s} = \frac{\partial \tilde{P_i}}{\partial z^{(i)}}\frac{\partial z^{(i)}}{\partial s}$. Now, we consider about $\tilde{P_{ij}}$ , the $j$-th component of $\tilde{P_{i}}$. By differentiating $\tilde{P_{ij}}$ with respect to ${z^{(i)}}_l$, we 
obtain 
\begin{equation*}
\begin{aligned}
\frac{\partial \hat{P_{ij}}}{\partial {z^{(i)}}_l} &= \begin{cases} \sigma ({z^{(i)}})_l \left(1 - \sigma ({z^{(i)}})_j \right) & \text{if } \ l=j,\\\\ - \ \sigma ({z^{(i)}})_j \cdot \sigma ({z^{(i)}})_l & \text{if } \ l \neq j. \end{cases}
\end{aligned}
\end{equation*}
where $\sigma(\cdot) = \text{softmax}(\cdot)$. Hence
\begin{equation*}
\begin{aligned}
\frac{\partial \hat{P_i}}{\partial z^{(i)}} %&= %\begin{bmatrix} \frac {\partial\hat{P_{i1}}}{\partial {z^{(i)}}_1} & \frac {\partial\hat{P_{i1}}}{\partial {z^{(i)}}_2} & \cdots & \frac {\partial\hat{P_{i1}}}{\partial {z^{(i)}}_n} \\ \frac {\partial\hat{P_{i2}}}{\partial {z^{(i)}}_1} & \frac {\partial\hat{P_{i2}}}{\partial {z^{(i)}}_2} & \cdots & \frac {\partial\hat{P_{i2}}}{\partial {z^{(i)}}_n} \\
%\vdots & \vdots & \ddots & \vdots \\ \frac {\partial\hat{P_{in}}}{\partial {z^{(i)}}_1} & \frac {\partial\hat{P_{in}}}{\partial {z^{(i)}}_2} & \cdots &\frac {\partial\hat{P_{in}}}{\partial {z^{(i)}}_n} \end{bmatrix} \\ 
&= \begin{bmatrix} \sigma({z^{(i)}})_1 \left(1 - \sigma ({z^{(i)}})_1 \right) & - \ \sigma ({z^{(i)}})_1 \cdot \sigma ({z^{(i)}})_2 & \cdots & - \ \sigma ({z^{(i)}})_1 \cdot \sigma ({z^{(i)}})_n \\ - \ \sigma ({z^{(i)}})_2 \cdot \sigma ({z^{(i)}})_1 & \sigma ({z^{(i)}})_2 \left(1 - \sigma ({z^{(i)}})_2 \right) & \cdots & - \ \sigma ({z^{(i)}})_2 \cdot \sigma ({z^{(i)}})_n\\
\vdots & \vdots & \ddots &\vdots \\- \ \sigma ({z^{(i)}})_n \cdot \sigma ({z^{(i)}})_1 & - \ \sigma ({z^{(i)}})_n \cdot \sigma ({z^{(i)}})_2 & \cdots & \sigma ({z^{(i)}})_n \left(1 - \sigma ({z^{(i)}})_n \right)\end{bmatrix}\\
&=\begin{bmatrix} \sigma ({z^{(i)}})_1 & 0 & \cdots & 0 \\ 0 & \sigma ({z^{(i)}})_2 & \cdots & 0\\
\vdots & \vdots & \ddots &\vdots \\0 & 0 & \cdots & \sigma ({z^{(i)}})_n \end{bmatrix} \\ & \qquad - \begin{bmatrix} \sigma ({z^{(i)}})_1 \cdot \sigma ({z^{(i)}})_1 & \sigma ({z^{(i)}})_1 \cdot \sigma ({z^{(i)}})_2 & \cdots & \sigma ({z^{(i)}})_1 \cdot \sigma ({z^{(i)}})_n \\ \sigma ({z^{(i)}})_2 \cdot \sigma ({z^{(i)}})_1 & \sigma ({z^{(i)}})_2 \cdot \sigma ({z^{(i)}})_2 & \cdots & \sigma ({z^{(i)}})_2 \cdot \sigma ({z^{(i)}})_n  \\
\vdots & \vdots & \ddots & \vdots \\ \sigma ({z^{(i)}})_n \cdot \sigma ({z^{(i)}})_1 & \sigma ({z^{(i)}})_n \cdot \sigma ({z^{(i)}})_2 & \cdots & \sigma ({z^{(i)}})_n \cdot \sigma ({z^{(i)}})_n \end{bmatrix}\\
&= \text{diag}(\sigma(z^{(i)})) - \sigma(z^{(i)})(\sigma(z^{(i)}))^T \\
&=: H^{(i)} \qquad(\ast)
\end{aligned}
\end{equation*}

\noindent $(v)$ From the definition of ${z^{(i)}}_j$, its partial derivative with respect to $s_l$ is
\begin{equation*}
    \begin{aligned}
    \frac{\partial {z^{(i)}}_j}{\partial s_l} &= \tau^{-1} \left((n+1-2i)\frac{\partial s_j}{\partial s_l} - \sum_{k;k \neq j} \text{sgn}_{jk}\frac{\partial}{\partial s_l} (s_j-s_k) \right)\\ &= \begin{cases} \tau^{-1} \left((n+1-2i) - \sum_{k;k \neq j} \text{sgn}_{jk} \right) & \text{if } \ l = j \\\\ \tau^{-1}\text{sgn}_{jl} & \text{if } \  l \neq j \end{cases}.
    \end{aligned}
\end{equation*}
Hence, 
\begin{equation*}
\begin{aligned}
    \frac{\partial z^{(i)}}{\partial s}
    &= \tau^{-1}\begin{bmatrix} n+1-2i & 0 & \cdots & 0 \\ 0 & n+1-2i & \cdots & 0\\
\vdots & \vdots & \ddots &\vdots \\0 & 0 & \cdots & n+1-2i \end{bmatrix}  \\ 
& \qquad + 
    \tau^{-1}\begin{bmatrix} - \sum_{k;k \neq 1} \text{sgn}_{1k} & 0 & \cdots & 0 \\ 0 & - \sum_{k;k \neq 2} \text{sgn}_{2k} & \cdots & 0\\
\vdots & \vdots & \ddots &\vdots \\0 & 0 & \cdots & - \sum_{k;k \neq n} \text{sgn}_{nk} \end{bmatrix}  \\ 
& \qquad + 
    \tau^{-1}\begin{bmatrix} \ 0 & \text{sgn}_{12} & \cdots & \text{sgn}_{1n} \\  \text{sgn}_{21} & 0 & \cdots & \text{sgn}_{2n} \\ \vdots & \vdots & \ddots & \vdots \\ \text{sgn}_{n1} & \text{sgn}_{n2} & \cdots & 0 \end{bmatrix} \\
    &= \tau^{-1} \left(\text{diag}((n+1-2i)\mathbf{1}) + \text{diag} \left(\left(-\sum_{k; k \neq j}\text{sgn}_{jk}\right)\mathbf{1} \right) + \begin{bmatrix} \text{sgn}_{jk} \end{bmatrix} \right) \\
    &=: \tau^{-1}(D^{(i)} + R) \qquad (\ast\ast)
\end{aligned}
\end{equation*}

By $(\ast)$ and $(\ast\ast)$, we obtain 
\begin{equation*}
    \frac{\partial \tilde{P_i}}{\partial s} = \tau^{-1}(H^{(i)} (D^{(i)} + R)).
\end{equation*}

Finally, from $(i) \sim (v)$, we derive
\begin{equation*}
\frac{\partial \neuloss}{\partial \mathbf{x}} = \frac{2}{\tau}\left(\tilde{P}_{[1:K]} - y_u \right)^T\sum_{i} (H^{(i)} (D^{(i)} + R)) \frac{\partial {\hat{y_u}}}{\partial \mathbf{x}}.
\end{equation*} 
Since $\nabla \neuloss = \left( \frac{\partial \neuloss}{\partial \mathbf{x}} \right)^T$, taking transpose on left- and right-hand side. Then we have
\begin{equation*}
\begin{aligned}
\nabla\neuloss 
&= \frac{2}{\tau} \left(\frac{\partial {\hat{y_u}}}{\partial \mathbf{x}} \right)^T \sum_{i} (H^{(i)} (D^{(i)} + R))^T \left(\tilde{P}_{[1:K]} - y_u \right)\\
&=: \frac{2}{\tau} \left(\frac{\partial {\hat{y_u}}}{\partial \mathbf{x}} \right)^T \sum_{i} W^{(i)} \left(\tilde{P}_{[1:K]} - y_u \right).
\end{aligned}
\end{equation*} 

%This completes the derivation. \qquad \qquad \qquad \qquad \qquad \quad \Box

Gradient.

For $\hingeloss$, 
\begin{equation} \label{eq:alpha-clr-update}
\frac{\partial \hat{y}_u}{\partial \mathbf{\uv}} = \begin{cases} \begin{bmatrix} \iv_1 \ \iv_2 \ \cdots \ \iv_n \end{bmatrix}^T & \text{score function is dot product} \\ -2\begin{bmatrix} \uv - \iv_1 \ \uv - \iv_2 \ \cdots \ \uv - \iv_n \end{bmatrix}^T & \text{score function is L2 distance} \end{cases},
\end{equation}
\begin{equation}\label{eq:beta-clr-update}
    \frac{\partial \hat{y}_u}{\partial \mathbf{\iv}_i} =  \begin{cases}
    \begin{bmatrix} \mathbf{0} \ \cdots \ \mathbf{0} \ \underbrace{\uv}_{{i \text{-th}}} \ \mathbf{0} \ \cdots \ \mathbf{0} \end{bmatrix}^T  & \text{score function is dot product} \\
    \begin{bmatrix} \mathbf{0} \ \cdots \ \mathbf{0} \ \underbrace{2(\uv - \iv)}_{{i \text{-th}}} \ \mathbf{0} \ \cdots \ \mathbf{0} \end{bmatrix}^T & \text{score function is L2 distance}   \end{cases},
\end{equation}\\
\begin{equation} \label{eq:alpha-hinge-update}
    \frac{\partial \hingeloss}{\partial \uv} = 
    \begin{cases} 
        \Phi_{ui} \indicator{[\gamma - s(u, i) + s(u, j) > 0]}(\iv_j - \iv_i)^T & \text{score function is dot product} \\
        2  \Phi_{ui} \indicator{[\gamma - s(u, i) + s(u, j) > 0]}(\iv_j - \iv_i)^T & \text{score function is L2 distance}
    \end{cases},
\end{equation} \\ 
\begin{equation} \label{eq:beta-hinge-update}
\begin{aligned}
    \frac{\partial \hingeloss}{\partial \iv_i} = 
    \begin{cases} 
        -\Phi_{ui} \indicator{[\gamma - s(u, i) + s(u, j) > 0]}{\uv_u}^T & \text{score function is dot product} \\
       2\Phi_{ui} \indicator{[\gamma - s(u, i) + s(u, j) > 0]}(\iv_i - \uv_u)^T & \text{score function is L2 distance}
    \end{cases}, \\ \\
    \frac{\partial \hingeloss}{\partial \iv_j} = 
    \begin{cases} 
       \Phi_{ui} \indicator{[\gamma - s(u, i) + s(u, j) > 0]}{\uv_u}^T & \text{score function is dot product} \\ 
       2\Phi_{ui} \indicator{[\gamma - s(u, i) + s(u, j) > 0]}(\uv_u - \iv_j)^T & \text{score function is L2 distance}
    \end{cases}.
\end{aligned}
\end{equation}

\maketitle

\section{Melon Dataset}
We give a detailed description of Melon dataset, which is not widely used in previous research works, and its description is written in Korean. This dataset is released for Melon playlist continuation contest, held on April 27th to July 26th, 2020 (https://arena.kakao.com/c/7/data). Melon is the largest music streaming service in South Korea.

The contest aims to learn models to predict appropriate songs that fit well with the given music playlist. 
The original data contains playlists and songs and their metadata. A playlist has several songs, and a single song can be in many playlists, establishing many-to-many relationship. 

Playlists can have tags. Tags can be shared among playlists, as songs are shared among playlists.
Although the dataset has rich metadata, such as titles of songs and playlists, tags, and \textit{likes} to the playlists and Mel-Spectrogram of the songs, we only take the membership information between songs and playlists as implicit feedback. 

We interpret a playlist as a user and songs in the dataset as items that the user has interacted with.  
Train and test datasets are released, but we use train dataset \texttt{train.json} as the whole dataset because the released test dataset is divided already into test input and test output.

\section{Update Rule} 
We have the gradient update rule $\nabla_{\hat{y}_u} \neuloss$ of $\neuloss$ with respect to a score vector $\hat{y}_u$:
\begin{equation} \label{eq:drm-update}
\begin{aligned}
\nabla_{\hat{y}_u} \neuloss = -\frac{2}{\tau} \sum_{k} W^{(k)} \left(y_u - \tilde{P}_{[1:K]}\right)
\end{aligned}
\end{equation}
where  
\begin{equation*}
\text{sgn}_{i, j} =\begin{cases} 1 & \text{if } \hat{y}_{u, i}>\hat{y}_{u, j} \\ 0 & \text{if } \hat{y}_{u, i}=\hat{y}_{u, j}\\-1 & \text{if } \hat{y}_{u, i}<\hat{y}_{u, j}\end{cases} \\
\end{equation*}
and
\begin{equation*}
\begin{aligned}
z^{(k)} &= \tau^{-1} \left((n+1-2k)\hat{y_u}-A_{\hat{y_u}} \mathbf{1} \right)\\ 
W^{(k)} &= w(k, K)(H^{(k)} (D^{(k)} + R))^T \\
H^{(k)} &= \begin{bmatrix} \sigma_1 E_1 \ \sigma_2 E_2 \ \cdots \ \sigma_n E_n \end{bmatrix} \\
& \ \ - \text{softmax}(z^{(k)})(\text{softmax}(z^{(k)}))^T \\ 
D^{(k)} &= (n+1-2k)I_n\\
R &= - \begin{bmatrix}  \sum_{l; \ l \neq 1} \text{sgn}_{1, l} E_1 \ \cdots \  \sum_{l; \ l \neq n} \text{sgn}_{n, l} E_n \end{bmatrix} \\
 & \ \ + \begin{bmatrix} \text{sgn}_{j, l} \end{bmatrix}_{n \times n}\\
\end{aligned}
\end{equation*}
with the $i$-th one-hot column vector $E_i \in \mathbb{R}^{n}$. \\

$\textit{Sketch of the derivation.}$ 

 $(i)$ Let $Q := \tilde{P}_{[1:K]}$ with $\tilde{P_k} = \begin{bmatrix} \tilde{P}_{k,1}, \tilde{P}_{k,2}, \cdots, \tilde{P}_{k,n}\end{bmatrix}^T$. \\For convenience, let $s=\hat{y}_u$ with $s_i = \hat{y}_{u,i}$, and $w^{(k)} := w(k, K)$. By Chain Rule, 
$$
\frac{\partial \neuloss}{\partial {s}} = \frac{\partial \neuloss}{\partial Q} \left( \sum_{k} \frac{\partial Q}{\partial \tilde{P_k}} \frac{\partial \tilde{P_k}}{\partial s}\right).
\qquad (\star)$$
Since $\neuloss = \|y_u - Q\|^2$, we can deduce 

\begin{equation*}
\begin{aligned}
\frac{\partial \neuloss}{\partial Q}
&= \frac{\partial}{\partial Q}\left( (y_u - Q)^T(y_u - Q) \right)\\
& = \frac{\partial}{\partial Q} \left( {y_u}^T {y_u} -Q^T y_u - {y_u}^T Q + Q^T Q \right) \\
&= -2(y_u - Q)^T .
\end{aligned}
\end{equation*}

 $(ii)$ By the definition of Q, we can derive 
\begin{equation*}
\begin{aligned}
\frac{\partial Q}{\partial \tilde{P}_k} &= \begin{bmatrix} \frac {\partial(\sum_l w^{(l)}\tilde{P}_{l1})}{\partial \tilde{P}_{k1}} & \frac {\partial(\sum_l w^{(l)}\tilde{P}_{l1})}{\partial \tilde{P}_{k2}} & \cdots & \frac {\partial(\sum_l w^{(l)}\tilde{P}_{l1})}{\partial \tilde{P}_{kn}} \\ \frac {\partial(\sum_l w^{(l)}\tilde{P}_{l2})}{\partial \tilde{P}_{k1}} & \frac {\partial(\sum_l w^{(l)}\tilde{P}_{l2})}{\partial \tilde{P}_{k2}} & \cdots & \frac {\partial(\sum_l w^{(l)}\tilde{P}_{l2})}{\partial \tilde{P}_{kn}} \\
\vdots & \vdots & \ddots & \vdots \\ \frac 
{\partial(\sum_l w^{(l)}\tilde{P}_{ln})}{\partial \tilde{P}_{k1}} & \frac {\partial(\sum_l w^{(l)}\tilde{P}_{ln})}{\partial \tilde{P}_{k2}} & \cdots & \frac {\partial(\sum_l w^{(l)}\tilde{P}_{ln})}{\partial \tilde{P}_{kn}} \end{bmatrix} \\ &= \begin{bmatrix} w^{(k)} & 0 & \cdots & 0 \\ 0 & w^{(k)} & \cdots & 0\\
\vdots & \vdots & \ddots &\vdots \\0 & 0 & \cdots & w^{(k)} \end{bmatrix}_{n \times n} \\ 
&= w^{(k)} I_n
\end{aligned}
\end{equation*}
for all $k=1,\ 2,\ 3,\ \cdots,\ n$.\\

 $(iii)$ Set $z^{(k)}:=\tau^{-1}((n+1-2k)s - A_s \mathbf{1}) \in \mathbb{R}^{n}$. With the property $|s_i - s_j| = \text{sgn}_{i, j}(s_i-s_j)$, we can write $A_s \mathbf{1}$ as
\begin{equation*}
(A_s \mathbf{1})_j=\sum_{l; \ l \neq j} \text{sgn}_{j, l}(s_j-s_l). 
\end{equation*}
Hence the $j$-th element ${z^{(k)}}_j$ of $z^{(k)}$ is 
\begin{equation*}
{z^{(k)}}_j=(n+1-2k)s_j - \sum_{l; \ l \neq j} \text{sgn}_{j, l}(s_j-s_l).
\end{equation*}\\
$(iv)$ Note that $\frac{\partial \tilde{P}_k}{\partial s} = \frac{\partial \tilde{P}_k}{\partial z^{(k)}}\frac{\partial z^{(k)}}{\partial s}$. Now, we consider about $\tilde{P}_{k, i}$ , the $i$-th element of $\tilde{P}_{k}$. Let $\sigma(\cdot) = \text{softmax}(\cdot)$. By differentiating $\tilde{P}_{k, i}$ with respect to ${z^{(k)}}_j$, we 
obtain 
\begin{equation*}
\begin{aligned}
\frac{\partial \tilde{P}_{k, i}}{\partial {z^{(k)}}_j} 
&= \frac{\partial}{\partial {z^{(k)}}_j} \left( \sigma(z^{(k)})_i \right) \\
&= \begin{cases} \sigma ({z^{(i)}})_j \left(1 - \sigma ({z^{(k)}})_i \right) & \text{if } \ j=i,\\\\ - \ \sigma ({z^{(k)}})_i \cdot \sigma ({z^{(k)}})_j & \text{if } \ j \neq i \end{cases} \\
&=: \begin{cases} \sigma_j \left(1 - \sigma_i \right) & \text{if } \ j=i,\\\\ - \ \sigma_i \cdot \sigma_j & \text{if } \ j \neq i. \end{cases}
\end{aligned}
\end{equation*}
Hence, with the $i$-th one-hot column vector $E_i$,
\begin{equation*}
\begin{aligned}
\frac{\partial \tilde{P}_k}{\partial z^{(k)}}  
&= \begin{bmatrix} \sigma_1 \left(1 - \sigma_1 \right) & - \ \sigma_1 \cdot \sigma_2 & \cdots & - \ \sigma_1 \cdot \sigma_n \\ - \ \sigma_2 \cdot \sigma_1 & \sigma_2 \left(1 - \sigma_2 \right) & \cdots & - \ \sigma_2 \cdot \sigma_n\\
\vdots & \vdots & \ddots &\vdots \\- \ \sigma_n \cdot \sigma_1 & - \ \sigma_n \cdot \sigma_2 & \cdots & \sigma_n \left(1 - \sigma_n \right)\end{bmatrix}\\
&=\begin{bmatrix} \sigma_1 & 0 & \cdots & 0 \\ 0 & \sigma_2 & \cdots & 0\\
\vdots & \vdots & \ddots &\vdots \\0 & 0 & \cdots & \sigma_n \end{bmatrix} \\ 
& \ \ - \begin{bmatrix} \sigma _1 \cdot \sigma_1 & \sigma_1 \cdot \sigma_2 & \cdots & \sigma_1 \cdot \sigma_n \\ \sigma_2 \cdot \sigma_1 & \sigma_2 \cdot \sigma_2 & \cdots & \sigma_2 \cdot \sigma_n  \\
\vdots & \vdots & \ddots & \vdots \\ \sigma_n \cdot \sigma_1 & \sigma_n \cdot \sigma_2 & \cdots & \sigma_n \cdot \sigma_n \end{bmatrix}\\
&= \begin{bmatrix} \sigma_1 E_1 \ \sigma_2 E_2 \ \cdots \ \sigma_n E_n \end{bmatrix} - \sigma(z^{(k)})(\sigma(z^{(k)}))^T \\
&=: H^{(k)} \qquad(\ast)
\end{aligned}
\end{equation*}

 $(v)$ From the definition of ${z^{(k)}}_i$, its partial derivative with respect to $s_j$ is
\begin{equation*}
    \begin{aligned}
    \frac{\partial {z^{(k)}}_i}{\partial s_j} &= \frac{1}{\tau} \left((n+1-2k)\frac{\partial s_i}{\partial s_j} - \sum_{l; \ l \neq i} \text{sgn}_{i, l}\frac{\partial}{\partial s_j} (s_i-s_l) \right)\\ &= \begin{cases} \frac{1}{\tau} \left((n+1-2k) - \sum_{l; \ l \neq i} \text{sgn}_{i, l} \right) & \text{if } \ j = i, \\\\ \frac{1}{\tau} \text{sgn}_{i, j} & \text{if } \  j \neq i. \end{cases}
    \end{aligned}
\end{equation*}
Hence, 
\begin{equation*}
\begin{aligned}
    \frac{\partial z^{(k)}}{\partial s}
    &= \frac{1}{\tau} \begin{bmatrix} n+1-2k & 0 & \cdots & 0 \\ 0 & n+1-2k & \cdots & 0\\
\vdots & \vdots & \ddots &\vdots \\0 & 0 & \cdots & n+1-2k \end{bmatrix}  \\ 
& - 
    \frac{1}{\tau} \begin{bmatrix} \sum_{l; \ l \neq 1} \text{sgn}_{1, l} & 0 & \cdots & 0 \\ 0 & \sum_{l; \ l \neq 2} \text{sgn}_{2, l} & \cdots & 0\\
\vdots & \vdots & \ddots &\vdots \\0 & 0 & \cdots & \sum_{l; \ l \neq n} \text{sgn}_{n, l} \end{bmatrix}  \\ 
& + 
    \frac{1}{\tau} \begin{bmatrix} \ 0 & \text{sgn}_{1, 2} & \cdots & \text{sgn}_{1, n} \\  \text{sgn}_{2, 1} & 0 & \cdots & \text{sgn}_{2, n} \\ \vdots & \vdots & \ddots & \vdots \\ \text{sgn}_{n, 1} & \text{sgn}_{n, 2} & \cdots & 0 \end{bmatrix} \\
    &= \frac{1}{\tau} (n+1-2k)I_n \\
    & \ \ - \frac{1}{\tau} (\begin{bmatrix}  \sum_{l; \ l \neq 1} \text{sgn}_{1, l} E_1 \ \cdots \  \sum_{l; \ l \neq n} \text{sgn}_{n, l} E_n \end{bmatrix} + \begin{bmatrix} \text{sgn}_{j, l} \end{bmatrix} ) \\
    &=: \frac{1}{\tau} (D^{(k)} + R) \qquad (\ast\ast)
\end{aligned}
\end{equation*}

By $(\ast)$ and $(\ast\ast)$, we obtain 
\begin{equation*}
    \frac{\partial \tilde{P}_k}{\partial s} = \frac{1}{\tau} (H^{(k)} (D^{(k)} + R)).
\end{equation*}\\

$(vi)$ Finally, from $(\star)$, we derive
\begin{equation*}
\frac{\partial \neuloss}{\partial {s}} = -\frac{2}{\tau}\left( y_u - Q \right)^T \sum_{k} w^{(k)}(H^{(k)} (D^{(k)} + R)).
\end{equation*} 
 Since $\nabla_s \neuloss = \left( \frac{\partial \neuloss}{\partial {s}} \right)^T$, from taking transpose on left- and right-hand side, we have
\begin{equation*}
\begin{aligned}
\nabla_s \neuloss 
&= - \frac{2}{\tau} \sum_{k} w^{(k)}(H^{(k)} (D^{(k)} + R))^T \left(y_u - Q \right)\\
&=: - \frac{2}{\tau} \sum_{k} W^{(k)} \left(y_u - Q \right).
\end{aligned}
\end{equation*} 
Therefore, 
\begin{equation*} \label{eq:drm-update}
\begin{aligned}
\nabla_{\hat{y}_u} \neuloss = -\frac{2}{\tau} \sum_{k} W^{(k)} \left(y_u - \tilde{P}_{[1:K]}\right).
\end{aligned}
\end{equation*}